\documentstyle[12pt]{article}
\newcommand{\be}{\begin{equation}}
\newcommand{\ee}{\end{equation}}
\newcommand{\ba}{\begin{eqnarray}}
\newcommand{\ea}{\end{eqnarray}}

\begin{document}
\begin{center}
{\bf\Huge  {Hamilton-Jacobi treatment of  fields with constraints
 }}
\end{center} \begin{center} Dumitru Baleanu\footnote{ Permanent address
: Institute of Space Sciences, P.O.BOX, MG-23, R 76900,Magurele-Bucharest,
Romania,E-Mail address:~~baleanu@thsun1.jinr.ru,baleanu@venus.nipne.ro}\\ Bogoliubov
Laboratory of Theoretical Physics \\ Joint Institute for Nuclear
Research\\ Dubna, Moscow Region, Russia

\end{center}
\begin{center}
    and
\end{center}
\begin{center}
 Yurdahan $G\ddot{u}ler$ \footnote{E-Mail address:~~yurdahan@ari.cankaya.edu.tr}
\end{center}
\begin{center}
Department of Mathematics and Computer Sciences ,Faculty of Arts and Sciences,
Cankaya University ,Turkey
\end{center}

\vskip 5mm
\bigskip
\nopagebreak
\begin{abstract}
In this paper  the $Guler's$ formalism for the systems with finite
degrees of freedom  is applied to the
field theories with constraints.
The integrability conditions are
investigated and the  path integral quantization is performed
using the action given by Hamilton-Jacobi formulation.
The Proca's model is investigated in details.
\end{abstract}

\section{Introduction}
The most common method for investigating the Hamiltonian treatment of
constrained systems was initiated by Dirac\cite{Dirac}.
The main feature of his method is to consider primary constraints
first.All  constraints are obtained using consistency conditions.
Hence , equations of motion are obtained  in terms of arbitrary
parameters.

The starting point of the $G\ddot{u}ler's$ method\cite{gu87},\cite{gu872},\cite{gu92},\cite{gu921},\cite{gu923},\cite{gu924},\cite{gu95}
is the variational
principle.
The Hamiltonian treatment of constrained systems leads us to total
differential equations in many variables.The equations are integrable
if the corresponding system of partial differential equations is a
Jacobi system.

Recently $G\ddot{u}ler$ has presented a treatment of classical fields
as constrained systems\cite{gu98}.
Then Hamilton-Jacobi quantization  of finite dimensional system
with constraints was investigated in \cite{balgu}.
 The purpose of this
paper is to generalize the $G\ddot{u}ler's$ formalism for systems
with finite degrees of freedom in order to include  the  field
theories with constraints also.

The plan of this paper is the
following:  In Sec.2 the $G\ddot{u}ler's$ formalism for
 the theories of fields
with constraints is presented  and the integrability conditions are
investigated.  In Sec.3 the quantization of Proca's model is analyzed.
In Sec.4 we present our conclusions.

\section{$G\ddot{u}ler's$ formalism for the field theories  with
constraints}
We generalize the $G\ddot{u}ler's$  formalism for systems finite dimensions
and constraints \cite{balgu} in order to describe the field theories
with constraints.  Let suppose that the local field theories are
described by the Lagrangean \be
L(\phi_{i},{\partial\phi_{i}\over\partial x_{\mu}}),i=1,\cdots,n
\ee

 The canonical formulation  gives the set of Hamilton-Jacobi
partial-differential equation as
\be
H_{\alpha}^{'}(\chi_{\beta},\phi_{\alpha},{\partial S\over\partial
\phi_{\alpha}},{\partial S\over\partial \chi_{\alpha}})=0,
\alpha,\beta=0,n-r+1,\cdots,n,a=1,\cdots,n-r,
\ee
where
\be\label{(ham)}
H_{\alpha}^{'}=H_{\alpha}(\chi_{\beta},\phi_{a},\pi_{a}) +\pi_{\alpha}
\ee
and $H_{0}$ is the canonical hamiltonian.
 The equations of motion are obtained as total differential equations
in many variables as follows
\be\label{(pq)}
d\phi_{a}={\partial H_{\alpha}^{'}\over\partial \pi_{a}}d\chi_{\alpha},
d\pi_{a}=-{\partial H_{\alpha}^{'}\over\partial \phi_{a}}d\chi_{\alpha},
d\pi_{\mu}=-{\partial H_{\alpha}^{'}\over\partial
\chi_{\mu}}d\chi_{\alpha}, \mu=1,\cdots, r \ee
\be\label{(z)}
dz=(-H_{\alpha} +\pi_{a}{\partial H_{\alpha}^{'}\over\partial
\pi_{a}})d\chi_{\alpha} \ee where $z=S(\chi_{\alpha},\phi_{a})$.The set
of equations(\ref{(pq)},\ref{(z)}) is integrable if \be\label{(h1)}
dH_{0}^{'}=0,dH_{\mu}^{'}=0,\mu=1,\cdots r
\ee
If conditions(\ref{(h1)}) are not satisfied identically , one considers
them as  new constraints and again tests the consistency conditions.
Thus repeating this procedure one may obtain a set of conditions.

\subsection{Integrability conditions}

If eqs.(\ref{(pq)}) are integrable , then the solutions of
eqs.(\ref{(z)}) are obtained by a quadrature.Hence, the investigation of
the integrability condition of eqs.(\ref{(pq)}) is sufficient.  As it was
discussed before in\cite{gu87} equations of motion of a singular system are
total differential equations.  As it is well known, to any set of total
differential equations
\be
dx_{i}=b_{i\alpha}(t_{\beta},x_{j})dt_{\alpha},
i,j=1,\cdots,n ,\alpha,\beta=0,1\cdots,p< n
\ee
there  are corresponding  set of partial differential equations in the
form \be b_{i\alpha}{\partial f\over\partial x_{i}}=0 \ee Now we may
investigate the integrability conditions of the equations (\ref{(pq)}).

To achieve this aim we  define the linear operators
$X_{\alpha}$ as \be X_{\alpha}f=b_{i\alpha}{\partial f\over\partial
x_{i}} \ee For the field theory we will define the linear operators
$X_{\alpha}$ as \be
X_{\alpha}f(\chi_{\beta},\phi_{a},\pi_{a})=[f,H^{'}_{\alpha}]
={\delta f\over\delta\phi_{a}}{\delta H_{\alpha}^{'}\over\delta\pi_{a}}
-{\delta f\over\delta\pi_{a}}{\delta H_{\alpha}^{'}\over\delta\phi_{a}}+
{\delta f\over\delta_{\chi_{\alpha}}}
\ee
{\bf{Lemma}}

A system of differential equations(\ref{(pq)})
is integrable iff
\be\label{(condh)}
[H_{\alpha}^{'},H_{\beta}^{'}]=0
\ee

{\bf Proof.}

 Suppose that (\ref{(condh)}) is satisfied.
Then
\be
(X_{\alpha},X_{\beta})=(X_{\alpha}X_{\beta}- X_{\beta}X_{\alpha})f
=X_{\alpha}[f,H_{\beta}^{'}]-X_{\beta}[f,H_{\alpha}^{'}]
\ee
we get after using Jacobi's identity
\be\label{(gen)}
 (X_{\alpha},X_{\beta})f=[f,[H_{\beta}^{'},H_{\alpha}^{'}]]
\ee

From(\ref{(condh)}) and (\ref{(gen)}) we conclude that
\be\label{(sis)}
(X_{\alpha},X_{\beta})=0
\ee
Conversely , if the system is complete , then(\ref{(sis)})
is satisfied for any $\alpha$ and $\beta$ and
we get
\be
[H_{\alpha}^{'},H_{\beta}^{'}]=0
\ee
Q.E.D.
\subsection{Quantization of  field theories with constraints}

In this section we will investigate the
quantization of the fields with constraints using $G\ddot{u}ler's$
formalism.

Let us suppose that  for a field with constraints we
found all independent hamiltonians $H_{\mu}^{'}$  using the
calculus of variations\cite{gu87},\cite{gu92},\cite{gu923}.At this
stage we can use Dirac's procedure of
quatization\cite{Dirac}.
We have
\be
H_{\mu}^{'}\Psi=0,\mu=0,n-r+1,\cdots,n
\ee
where $\Psi$ is the wave
function.
The consistency conditions are
\be\label{(cond)}
[H_{\mu}^{'},H_{\nu}^{'}]\Psi=0,\mu,\nu=1,\cdots r
\ee
where $[,]$ is the commutator.
If  the hamiltonians $H_{\mu}^{'}$
satisfy
\be
[H_{\mu}^{'},H_{\nu}^{'}]=C_{\mu\nu}^{\alpha}H_{\alpha}^{'}
\ee
then we have a theory with a  first class constraints.

In the case when the hamiltonians $H_{\mu}^{'}$  satisfy
\be
[H_{\mu}^{'},H_{\nu}^{'}]=C_{\mu\nu}
\ee
with $C_{\mu\nu}$  not depending  on $\phi_{i}$ and $\pi_{i}$,
  from(\ref{(cond)})
there  arises naturally  Dirac's brackets and the canonical
quatization will be performed taking Dirac's brackets into
commutators.

 On the other hand  $G\ddot{u}ler's$ formalism gives an action when
all hamiltonians $H_{\mu}^{'}$  are in involution.  Because in
$G\ddot{u}ler's$ formalism we work from the beginning in the extended
space we suppose that variables $\chi_{\alpha}$ depend of $\tau$.Here
$\tau$ is canonical conjugate with $p_{0}$.

If we are able , for a given
 system with constraints, to find the independent hamiltonians
$H_{\mu}^{'}$ in involution then we can perform the quantization of
this system using the path integral quantization method with the action
given by(\ref{(z)}).
After some calculations we found that the action z has the following
form

\be\label{(z2)}
z=\int(-H_{\alpha} +\pi_{a}{\partial H_{\alpha}^{'}\over\partial
\pi_{a}})\dot{\chi_{\alpha}}d{\tau}
\ee
where
$\dot\chi_{\alpha}={d\chi_{\alpha}\over d\tau}$.
\section{The Proca's model}

The Proca's model is described by the Lagrangian
\be
L=-{1\over 4}F_{\mu\nu}F^{\mu\nu} + {m^{2}\over 2}A^{\mu}A_{\mu}
\ee
The form of the hamiltonian in $G\ddot{u}ler's$ formalism is:
\be
H_{0}^{'}= p_{0} +\int\left[{\pi^{2}\over
2} + {1\over 4}F_{ij}^{2} +{m^{2}\over 2}\left(A_{0}^{2}
+A_{i}^{2}\right)^{2} -A_{0}\phi_{2}\right]d^{3}x
\ee
 where
\be
\phi_{2}=\partial_{i}\pi^{i} +m^{2}A_{0}
\ee
The system posseses the primary constraint
\be
 H_{1}^{'}=\pi_{0}
\ee
Imposing
\be
dH_{0}^{'}=0
\ee
we get another constraint
\be
H_{2}=\partial_{i}\pi^{i} +m^{2}A_{0}
\ee

Then in the $G\ddot{u}ler's$ formalism we have three hamiltonians.
The hamiltonian $H_{2}$ is not yet in the form (\ref{(ham)}).
Since
\be\label{(h)}
[H_{1}^{'},H_{2}]=-m^{2}\delta\left(x-y\right)
\ee
the hamiltonians are  not in involution.
At this stage we can investigate the canonical quantization method
using Dirac's formalism.
From (\ref{(h)}) we conclude that the system have two second class
constraints in Dirac's classification and for canonical quantization we
need Dirac's brackets
\be
\{F,G\}_{D.B.}=\{F,G\} -\{F,H_{2}^{'}\}C^{21}\{H_{1}^{'},G\}
-\{F,H_{1}^{'}\}C^{12}\{H_{2}^{'},G\}
\ee
where

$\{,\}$ are the Poisson-brackets and the matrix $C^{\alpha\beta}$ is the
inverse of the matrix
\ba
C_{\alpha\beta}=\pmatrix{&0& -m^{2}\delta(x-y)\cr
		& m^{2}\delta(x-y)&0}
\ea

On the other hand in the $G\ddot{u}ler's$ formalism we have an action
which is well defined when the hamiltonians are in involution.  In our
case we found the hamiltonians
$\tau_{0}^{'}$,$\tau_{1}^{'}$,$\tau_{2}^{'}$ in involution in the
following form \ba \tau_{1}^{'}=H^{'}_{1} +m^{2}\rho=\pi_{0}
+\tau_{1}\cr \tau_{2}^{'}=H_{2} + \pi_{\rho}=\pi_{\rho}+ \tau_{2}\cr
\tau_{0}^{'}=p_{0} + H^{(1)} + H^{(2)}= p_{0} +\tau_{0} \ea where \be
H^{(1)}=\int d^{3}x \left[\left(\partial_{i}A^{i}\right)m^{2}\rho
-{\pi_{\rho}\over m^{2}}\left(\partial_{i}\pi^{i}
+m^{2}A_{0}\right)\right] \ee \be H^{(2)} =\int d^{3}x \left[{-1\over
 2m^{2}}\pi_{\rho}^{2} -{m^{2}\over
 2}\left(\partial_{i}\rho\right)\left(\partial^{i}\rho\right)\right]
\ee
and
 $\rho,\pi_{\rho}$
are the extra fields satisfying $\{\rho,\pi_{\rho}\}=1$ all the
other commutation relations become zero.

The action z has the following form
\be
dz=[-\tau_{0} +\int(\pi_{i}^{2}+
{\partial_{i}A_{0}}+{{\partial_{i}\pi{\rho}\over m^{2}})]d^{3}x
d{\tau}+ \int(-m^{2}\rho})d^{3}x dA_{0} +\int(-{\partial_{i}\pi^{i}}
+m^{2}A_{0})d^{3}x d{\rho}
\ee
or
\be\label{(Proca)}
z=\int{d\tau d^{3}x}\{[-\tau_{0} +\int(\pi_{i}^{2}+
{\partial_{i}A_{0}}+{\partial_{i}\pi{\rho}\over m^{2}})]
+ \int(-m^{2}\rho){\dot {A_{0}}} +\int(-{\partial_{i}\pi^{i}}
+m^{2}A_{0})\dot{{\rho}}\}
\ee
Here ${\dot A_{0}}={dA_{0}\over d\tau}$ and
${\dot\rho}={d\rho\over d\tau}$.

{ } For a system  with r first class-constraints $\psi^{\alpha}$
the path integral representation is given as\cite{senj}
\be\label{(int1)} <\phi^{'}\mid exp[-i(t^{'}-t){\hat H_{0}}]\mid
\phi>=\int\prod
d\mu(\phi_{\mu},\pi_{\mu})exp[i\{\int_{-\infty}^{+\infty}dt(\pi_{\mu}{\dot\phi_{\mu}}-H_{0})\}]
\ee
where the measure of integration is given as
\be
d\mu(\phi,\pi)=det\mid\{\psi^{\alpha},\psi^{\beta}\}\mid\prod
\delta(\chi^{\alpha})\delta(\phi^{\alpha}) \prod d\phi^{\mu}d\pi_{\mu}
\ee
and $\psi^{\alpha}$ are r- gauge constraints.

We found after some calculations that the action(\ref{(Proca)}) give us
the same result as(\ref{(int1)}) for the Proca's model when $\tau=t$
but using different gauge conditions.

\section{Concluding remarks}
 $G\ddot{u}ler$ has
initiated a new formalism for quantization of systems with constraints
\cite{gu87},\cite{gu872},\cite{gu92},\cite{gu921},\cite{gu923},\cite{gu924},\cite{gu95}.
 In this paper we have generalized the $G\ddot{u}ler's$ formalism
for the system with finite degrees of freedom\cite{balgu} to
 the  field  theories with constraints.Our formalism is completely
different from the formalism presented in\cite{gu98}.

 An interesting case appears  when  a theory has secondary
 constraints and the constraints are of second class in the Dirac's
classification.
We found that for the system with second class constraints
the Dirac's brackets arise naturally in the process of  quantization
for  field with constraints in the $G\ddot{u}ler's$ formalism.
The Dirac's brackets are defined on the extended space.

When the system with constraints has only primary constraints in involution
$G\ddot{u}ler's$ formalism give us exactly
the action  which has the same expression as
obtained in the path integral quantization after performing
all calculations.In this case we do not need any gauge conditions.
 If the system has secondary
constraints or second class constraints this result is not valid
since the hamitonians are not in involution.To obtain the
system in involution we need to extend the system.
Because in the $G\ddot{u}ler's $ formalism we have the freedom to
choose the dependence of gauge variables $\chi_{\alpha}$
we will choose $\chi_{\alpha}=\chi_{\alpha}(\tau)$.
In this case the action of $G\ddot{u}ler's$ formalism gives the same
results as  path integral formulation for the system with constraints.
 For the
Proca's model we extend the system and we found three hamiltonians in
involution.  The path integral quantization was performed with an
action given by $G\ddot{u}ler's$ formalism and the results are in
agreement with those obtained by other methods.

\section{Acknowledgements}

One of the authors (D.B.) would like to thank TUBITAK and NATO for financial
support  and METU for the hospitality during his
working stage at Department of Physics.

\end{document}